\documentclass[aps,prb,showpacs,twocolumn,groupedaddress]{revtex4-1}

\usepackage{graphicx,graphics,color,epsfig}
\usepackage{amsmath}
\usepackage{amssymb}
\usepackage{amsfonts}
\usepackage{amsfonts}
\usepackage{epstopdf}
\usepackage{bm}
\usepackage{times,xspace}
\usepackage{color}

\begin{document}


\title{Weyl semimetals and superconductors designed in an orbital selective superlattice}

\author{Tanmoy Das}
\affiliation{Theoretical Division, Los Alamos National Laboratory, Los Alamos, New Mexico 87545 USA}

\date{\today}

\begin{abstract}
We propose two complementary design principles for engineering three-dimensional (3D) Weyl semimetals and superconductors in a layer-by-layer setup which includes even and odd parity orbitals in alternating layers $-$ dubbed an orbital selective superlattice. Such a structure breaks mirror symmetry along the superlattice growth axis which, with the help of either a basal plane spin-orbit coupling or spinless $p+ip$ superconductivity, stabilizes a 3D Dirac node. To explore this idea, we develop a 3D generalization of the Haldane model and a Bogoliubov-de-Gennes (BdG) Hamiltonian for the two cases, respectively, and show that tunable single or multiple Weyl nodes with linear dispersion in all spatial directions can be engineered desirably in a widespread parameter space. We also demonstrate that a single helical Weyl band can be created at the $\Gamma$-point at the Fermi level in the superconducting case via gapping out either of the orbital states by violating its particle-hole symmetry but not any other symmetries. Finally, implications of our results for the realization of an anomalous Hall effect and Majorana bound state are discussed. 
\end{abstract}

\maketitle

\section{Introduction}

The Weyl semimetal is a new class of three-dimensional (3D) Dirac materials in which gapless Dirac excitations form at the linear contact of two nondegenerate bands at points or lines, and disperse in all spatial dimensions.\cite{Weyl1,Weyl2,WeylIridates,AHE} These Weyl points or nodes are essentially a product of a two-dimensional (2D) Dirac point split by either time-reversal ($\mathcal{T}$)\cite{WeylIridates,WeylTIBalentsTRB,WeylPGS} or inversion ($\mathcal{I}$)\cite{WeylTIBalentsISB,WeylKane,Ojanen} symmetry breaking; yet they remain topologically protected owing to other invariant symmetry(s) such as the space group,\cite{WeylKane} point group symmetry\cite{WeylPGS}, or lattice symmetry\cite{WeylFCC}. The topological protection of the Weyl points is quantified by a vanishing net Chern number (but individually the Chern number is an integer at each Weyl node), evaluated inside a 3D Berry curvature surrounding each valence Weyl point. Two such topologically protected Weyl points of opposite Chern number, separated in space by $\mathcal{I}$ symmetry, act as monopoles.\cite{WeylReviewVishwanath} Other interesting implications of Weyl semimetals include the realization of the anomalous Hall effect,\cite{AHE,WeylTIBalentsTRB,HgCr2Se4} nontrivial electromagnetic responses,\cite{WeylReviewVishwanath,axion,MFranz} Majorana fermions,\cite{WeylBalentsSC} and a disconnected$-$yet protected$-$Fermi surface or a so-called `Fermi arc'.\cite{WeylIridates,WeylReviewVishwanath} Furthermore, the Weyl excitations are protected from impurity scattering with a shorter scattering length than the 2D Dirac fermions,\cite{WeylImpurity} and thus open new avenues for potential applications and technologies.

A possible material realization of Weyl excitations has been put forward recently in several condensed matter systems. In a seminal work, Wan {\it et al.}\cite{WeylIridates} have proposed that Weyl nodes exist in pyrochlore iridates due to an interplay of spin-orbit coupling (SOC) and electronic interaction. Further postulates include a multilayer of a magnetically doped 3D topological insulator and a normal insulator,\cite{WeylTIBalentsTRB} ferromagnetic HgCr$_2$Se$_4$ spinel,\cite{HgCr2Se4} crystalline $A_3$Bi ($A=$Na, K, Rb),\cite{A3Bi} and $\beta$-cristobaline BiO$_2$.\cite{WeylKane} Despite these extensive theoretical efforts, experimental confirmation of its existence is still lacking, with the exception of an indirect hint of a band touching of the valence and conducting bands at the phase transition point between a non-trivial and trivial 3D topological insulator in BiTl(S$_{1-\delta}$Se$_{\delta}$)$_2$.\cite{WeylTI,ZHasan,BiTlSeStheory}

Here we go beyond the earlier proposals, and propose that Weyl nodes can also be easily stabilized by orbital symmetry in a layer-by-layer approach available within the molecular beam epitaxy (MBE) technique. We formulate two complementary design principles for this mechanism in an orbital selective layered structure with the inclusion of either SOC, or a spinless $p+ip$-wave superconducting (SC) state. The term orbital selective layer stands for a superlattice structure in which an even parity orbital layer is sandwiched in between two layers of an odd parity orbital. We find that in both cases, single or multiple Weyl nodes can be engineered and planted in various locations in the momentum space by taking advantage of the flexibility brought out in a heterostrucutre geometry. We also find that a single and nondegenerate Weyl point can be produced at the $\Gamma$-point in the SC state by breaking the particle-hole symmetry in only one of the layers in the orbital selective superlattice. 

We arrange the rest of the paper as follows. In Sec.~\ref{Sec:GC}, we layout the general concept of the orbital selective superlattice and how it can be used to stabilize the Weyl node(s). The Weyl Hamiltonian based on a tight-binding model is deduced in Sec.~\ref{Sec:WSOC} using basal-plane Rashba-type SOC and a spontaneous magnetization. The same Hamiltonian using $p+ip$-pairing symmetry is derived in Sec.~\ref{Sec:WSC} and the limiting case of obtaining a non-degenerate Weyl node at the $\Gamma$-point is discussed here. Finally we conclude in Sec.~\ref{Sec:Conclusion}. 

\section{General layout of orbital selective superlattice}\label{Sec:GC}

\begin{figure}[top]
\rotatebox[origin=c]{0}{\includegraphics[width=.9\columnwidth]{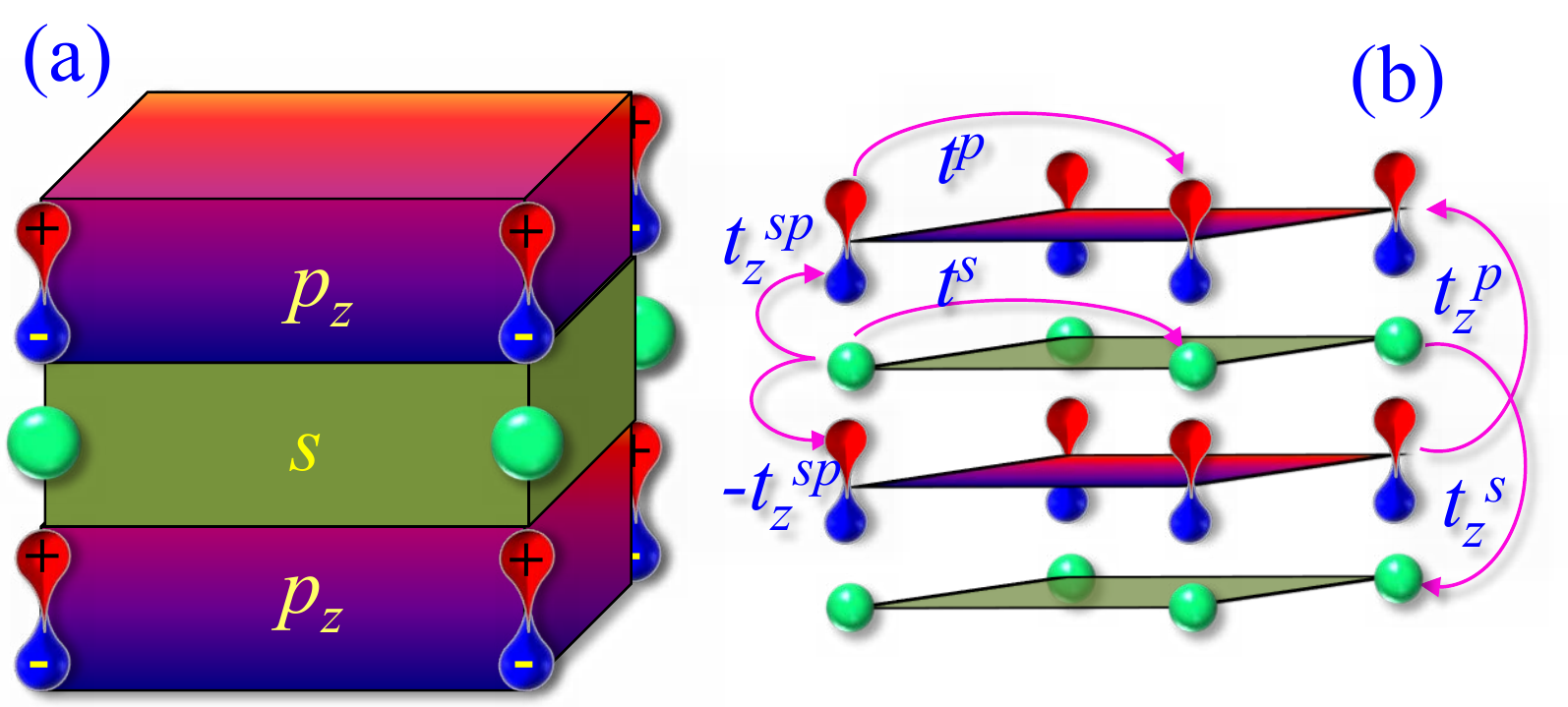}}
\caption{(Color online) (a) Proposed orbital-selective superlattice (schematic). The top and bottom layers inhabiting $p_z$-orbital symmetry are to be grown on both sides of an even parity $s$-orbital layer. An inter-orbital SOC or spinless chiral $p+ip$ SC is assumed to be present on the heterostructure. (b) Various tight binding hopping parameters are depicted for such an orbital-selective multilayer. The key to notice here is the change in sign in $t^{sp}_z$ for hopping on both sides of the $s$-layer, which breaks the mirror symmetry along the heterostructure growth axis. However, the spin-rotational symmetry breaking for the case of SOC, or the particle-hole symmetry for the SC case, restores the time-reversal invariance in this heterostucture. 
}\label{fig1}
\end{figure}

The general form of the Hamiltonian can be deduced from the starting point of ${\bm k}\cdot{\bm p}$ theory which takes the form of ${\bm d}\cdot{\bm \Gamma}$ for the 3D case, where ${\bm \Gamma}$ are the $4\times 4$ Dirac matrices, and the corresponding ${\bm d}$ vector represents a given structure. The orbital symmetry and superlattice structure can be progressively engineered together in such a way that each component ${\bm d}_i$ complements its corresponding $\Gamma_i$-matrix to lead to det$[{\bm d}_i]\ne0$ to stabilize a protected Weyl point. $\Gamma_{1,2}=\tau_x\otimes\sigma_{x,y}$ (${\bm \tau}$ and ${\bm \sigma}$ are the Pauli matrices in the orbital and spin basis, respectively) are the projection matrices aligned in the basal plane (say the $x$-$y$-plane) of each layer. $\Gamma_1$ and $\Gamma_2$ can thus be interlocked via a chiral SOC or chiral $p+ip$-wave SC. The chirality of these states manifests into linearly dispersive Dirac excitations in the basal plane. For the stability of these Dirac points along the $z$-axis, we consider the $\Gamma_3=\tau_x\otimes\sigma_z$ term. Here we propose an orbital-selective superlattice in which we take the $p_z$-orbital for the odd-parity layers, and the $s$-orbital on the even parity layer, as depicted in Fig.~\ref{fig1}(a). This setup clearly breaks the mirror ($\mathcal{M}$) symmetry, defined as $\mathcal{M}=-\mathbb{I}\otimes i\sigma_z$, with respect to the $s$-mirror plane. However, to gain a time-reversal invariance of the Hamiltonian, it is required to compensate for the broken $\mathcal{M}$ symmetry. This we achieve by employing a spin-polarized ground state with spin polarization aligned with the mirror symmetry axis (such as he magnetic impurity or spin-exchange term) which violates the spin-rotational symmetry $\mathcal{R}=-\mathbb{I}\otimes \sigma_x\mathcal{K}$ where $\mathcal{K}$ is the complex conjugation. A byproduct of these symmetry considerations is that $\mathcal{M}$ and $\mathcal{R}$ commute with each other, and their product gives the time-reversal operator $\mathcal{T}=\mathbb{I}\otimes i\sigma_y\mathcal{K}$. The Hamiltonian, thereby, becomes $\mathcal{T}$ invariant.

For a SOC locked 2D layer, we can think of a Bi-film,\cite{BiRashba} Pb/Ge surface\cite{PbRashba} or gated silicene film\cite{SiRashba} on our $p$-layer, which is demonstrated to inherit Rashba-type SOC. The added benefit would be that the SOC brings out the $p_z$ orbital at the Fermi level in these systems which serves our purpose. For the spinless SC case, the particle-hole symmetry ($\mathcal{S}$) does the job for the broken $\mathcal{R}$ symmetry case and the resulting Hamiltonian also remains $\mathcal{T}$ invariant. For such proximity induced $p+ip$-wave Cooper pairing on both layers, we can think of a substrate made of $^3$He A phase,\cite{He3} Sr$_2$RuO$_4$,\cite{Sr2RuO4} the heavy-fermion system CePt$_3$Si,\cite{CePt3Si} or an oxide interface.\cite{ointerface} The additional Dirac matrix $\Gamma_4=\tau_z\otimes\mathbb{I}$, which usually arises in a multi-orbital framework or due to the involvement of any gap parameter, will be used as a `tuning knob' to manipulate the $\mathcal{I}$-symmetry, and thereby to plant the Weyl points at our desired locations in the 3D momentum space.

\section{SOC induced Weyl nodes}\label{Sec:WSOC} 

Based on the aforementioned design method, we write down a generalized tight-binding Hamiltonian for the orbital selective superlattice with SOC as shown in Fig.~\ref{fig1}(b).
\begin{eqnarray}
H&=&-\sum_{j,\sigma,n\ne m}\Big[\mu^nc^{n\dag}_{j\sigma}c^n_{j\sigma}+
\frac{t_{j}^n}{2}c^{n\dag}_{j\sigma}c^n_{j\pm1\sigma} - i\nu\sigma \frac{t_z^{nm}}{2}c^{n\dag}_{j\sigma}c^m_{j\pm1\sigma} \nonumber\\
&&~~~~~~~~~~~~~ - i\frac{\alpha^{nm}}{2} c^{n\dag}_{j\sigma}\big({\bm \sigma}\times{\hat{\bm d}}_{nm}\big)\cdot{\bm z}c^m_{j\pm1\bar{\sigma}}  \Big].
\end{eqnarray}
Here $c^n_{j\sigma}~(c^{n\dag}_{j\sigma})$ is the creation (annihilation) fermionic operator at a lattice site $j$, and $n,m$ are orbital indices. $\nu=\pm$ is the broken $\mathcal{M}$ symmetry index with respect to the $s$-layer [as shown in Fig.~\ref{fig1}(b)], and $\sigma=\pm$ is the spin index with the easy axis aligned with that of $\nu$. The first term is the chemical potential for different orbitals, while the second term is an intraorbital nearest neighbor hopping along all three spatial directions as depicted in Fig.~\ref{fig1}(b). The intraorbital hopping integral along the superlattice growth axis, $t_z^n$, is the next-nearest neighbor term in this geometry, which is usually weaker in amplitude than the in-plane ones, and easily tunable. The third term is the spin-dependent interorbital hopping, which clearly breaks the spin-rotational ($\mathcal{R}$) symmetry ($\sigma=\pm$). However, as mentioned earlier, the $\mathcal{R}$ symmetry index ($\sigma$) and $z\rightarrow -z$ $\mathcal{M}$ symmetry index $\nu=\pm$ compensate each other. The last term is the SOC effect, introduced by Haldane\cite{Haldene} and later by Kane-Mele,\cite{KaneMele} which naturally commences a nearest neighbor Rashba-type interaction between the $s$ and $p$-orbitals due to an already broken $\mathcal{M}$ symmetry. $\hat{\bm d}_{nm}$ is the unit vector that connects the nearest neighbor sites between different layers.\cite{footnote}

To express the Hamiltonian in the usual $\Gamma$-matrix form, we use a basis of four-component `orbital-spinors' $\Psi_{\bm k}=(c^s_{{\bm k}\uparrow}, c^s_{{\bm k}\downarrow}, c^p_{{\bm k}\uparrow}, c^p_{{\bm k}\downarrow})$ of Bloch states with a 3D periodic boundary condition. However, it is easy to realize that a 2D electron gas with opposite parity on adjacent layers and open boundary conditions would also lead to the same low-energy effective Hamiltonian, and produce Weyl nodes of the same nature. The Fourier transformation of $H$ on a 2D square lattice gives $H=\Psi_{\bm k}^{\dag}H_{\bm k}\Psi_{\bm k}$, where
\begin{equation}\label{E:Hk}
H_{\bm k} =  \xi_{\bm k}^+\mathbb{I}+ \xi_{\bm k}^-\Gamma_4 +\alpha(s_{x}\Gamma_1+s_{y}\Gamma_2) -t^{sp}_z s_{z/2}\Gamma_3,
\end{equation}
where $\xi_{\bm k}^{\pm}=(\xi_{\bm k}^{s}\pm\xi_{\bm k}^p)/2$, with $\xi_{\bm k}^{s,p}=-t^{s,p}(c_x+c_y)-t_z^{s,p}c_z-\mu^{s,p}$, $c_{i} (s_{i}) =\cos{k_i} (\sin{k_i})$ and $i=x,y,z$. The energy spectrum of $H_{\bm k}$ is $E^{\pm}_{\bm k}=\xi_{\bm k}^+\pm\sqrt{(\xi_{\bm k}^-)^2+\alpha^2(s^2_x+s^2_y)+(t^{sp}_z)^2s_{z/2}^2}$. Notice that along the $z$-axis, the nearest-neighbor interorbital hopping $t^{sp}_z$ covers half of the distance that the next nearest-neighbor intra-orbital hoppings $t_z^{s,p}$ do, and that yields a factor of half in the former phase factor $s_{z/2}=\sin{(k_z/2)}$. For the case of preserved $\mathcal{T}$ and ${\mathcal I}$ symmetries, each band is doubly degenerate, and the Weyl points can be planted at the locus of $\xi^-_{{\bm k}}=0$ in the 3D space. Clearly the mechanism of gapless Weyl nodes does not rely on other parameters.

\begin{figure}[top]
\rotatebox[origin=c]{0}{\includegraphics[width=.9\columnwidth]{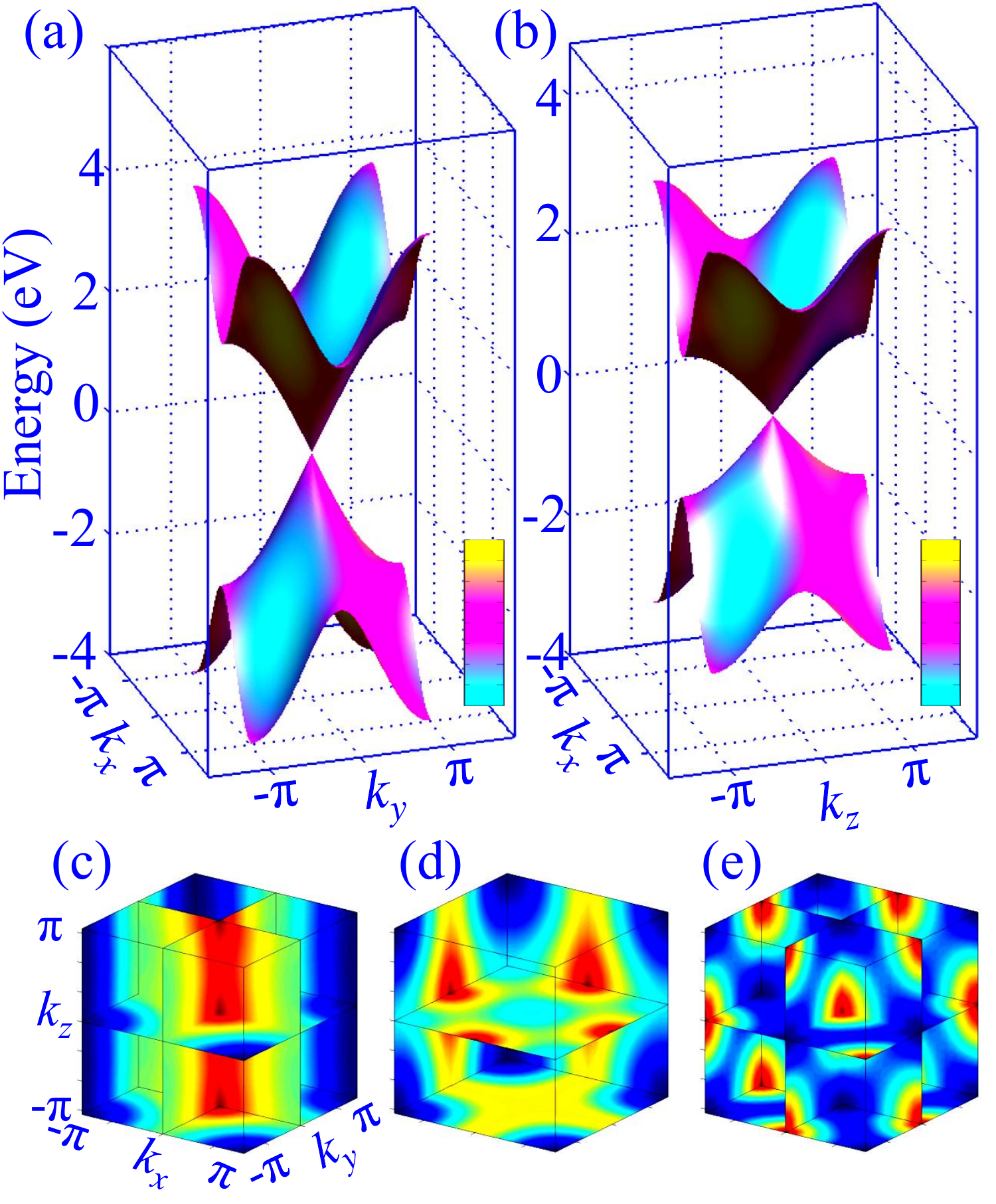}}
\caption{(Color online) (a) Computed band structure for the Hamiltonian in Eq.~(\ref{E:Hk}) with the presence of SOC. Here the results are projected on the basal plane in which the spin-chirality due to the SOC term [third term in Eq.~(\ref{E:Hk})] is shown in a colormap. (b) Same (a) but on the $x$-$z$-plane. The color gradient along the $x$ axis depicts spin orientation due to $\sigma_x$, while that along the $z$ direction is due to orbital polarization embedded in $\sigma_z$. (c),(d) Dispersion spectra in a blue to red colormap in the 3D spatial dimensions are presented for three representative cases of 3, 4 and 13 Weyl nodes (see text). The red color gives the locus of 3D Weyl points on the high-symmetry momenta. The same results are obtained for the BdG Hamiltonian in Eq.~(\ref{HSC}), in which the in-plane spin-polarization should be read as the $p+ip$ chirality (not shown). }\label{fig2}
\end{figure}

The flexibility of this design principle allows us to obtain a single Weyl node at the $(E,{\bm k})=(0,0)$ point for a tunable chemical potential of $\mu^{s,p}=-2t^{s,p}-t^{s,p}_z$. It is noteworthy that the difference $\mu^s-\mu^p$ is the most important tuning factor to close the gap at any momentum point(s) or lines, while the $\mu^s+\mu^p$ combination merely elevates the Weyl node to the Fermi level. Sample results are shown for a simplified and realistic parameter choice: $t^p=-1,~\alpha=1,~t_z^s=t_z^p=t_z^{sp}=0.5$ in units of $t^s=1$eV. The corresponding band dispersion in the 2D planes at $k_z=0$ and $k_y=0$ cuts are shown in Figs.~\ref{fig2}(a) and \ref{fig2}(b), with a colormap representing the spin polarization $\sigma_{x,y}$ and  $\sigma_{x,z}$, respectively. The Weyl fermion velocity on the basal plane is determined by the SOC strength $\alpha$, while that along the superlattice growth axis is controlled by the hopping $t_z^{sp}$ term. The density of states for the 3D Weyl state is quadratic in energy around the Fermi level, despite the presence of linear dispersion.

\subsection{Anomalous Hall effect} 

The anomalous Hall effect (AHE) has been proposed to exist in the Weyl semimetal under pressure,\cite{AHE} or the $\mathcal{T}$-symmetry breaking.\cite{WeylTIBalentsTRB,HgCr2Se4} For our geometry, an additional `knob' for generating AHE emanates from an externally triggered spatial potential with respect to the $s$-layer. To see this, we use the generalized formula for the Chern number in (3+1) dimensions in terms of the Green's functions as\cite{ChernGreen}
\begin{equation}
C=\frac{\pi}{6}\int_{\Lambda}\frac{d{\bm k}d\omega}{(2\pi)^4}{\rm Tr}\left[\epsilon^{\mu\nu\eta\gamma}\prod_{i=\mu\nu\eta\gamma}G\partial_iG^{-1}\right],
\end{equation}
where $\epsilon^{\mu\nu\eta\gamma}$ is the Levi-Civita symbol which runs over 3D spatial and energy dimensions. The cutoff scale $\Lambda$ restricts the integration range to the linear dispersion region, where the Green's function can be defined as $G({\bm k},\omega)=(\omega-{\bm d}\cdot{\Gamma})/(\omega^2-\sum_id_i^2)$, with the components of the $d$ vector being the linear expansion of  Eq.~(\ref{E:Hk}). The $C_4$-symmetry invariance of each orbital plane manifests into a vanishing net Chern number at any $k_z$ cut. A finite value of $C$ can be easily governed in our setup by playing with the ${\Gamma_3}$ term. This can be done by introducing a spatial gradient field or spontaneous magnetization along the $z$ axis. For such a case, the above formula converges to a total basal plane AHE $\sigma_{xy}=\pm\frac{e^2}{h}\big(\Delta m_z\Delta t^{sp}_z/t_{z0}^{sp}\big)/2$, where $\Delta t_z^{sp}$ is the hopping anisotropy induced by the spatial gradient with respect to its mean value $t_{z0}^{sp}$, and $\Delta m_z$ is any associated `spontaneous' magnetization. An interesting property known in different contexts is that the resultant electric polarization $P_z$ for such a case couples nonlinearly to the external electromagnetic field combination ${\bm E}\cdot{\bm B}$, and thereby may render non-trivial properties such as charge fractionalization or a magnetoelectric effect.\cite{electricP}

\subsection{Weyl node relocation}

 The heterostructure geometry also brings out an opportunity to plant the Weyl nodes at desired points or lines, by lifting the $\mathcal{I}$ symmetry in $\Gamma_4$ term. Some representative examples are shown in Figs.~\ref{fig2}(c)-\ref{fig2}(e). By switching off the intra-orbitals hopping along the heterostructure growth axis, i.e. $t_z^p=t_z^s=0$, we obtain an odd number of Weyl points at the high-symmetry momenta ${\bm k}$=(0,0,$\pm$1), in addition to one at the $(0,0,0)$ point; see Fig.~\ref{fig2}(c). Again, for $\mu^{s,p}=-t_z^{s,p}$, one can engineer Weyl nodes at ${\bm k}$=($\pm$1,0,0), (0,$\pm$1,0), and its equivalent points$-$an even number of Weyl nodes in total$-$as shown in Fig.~\ref{fig2}(d). Furthermore, by extending the electron overlap integrals to the next nearest-neighbor hopping, which yields $\xi_{\bm k}^{s,p}=-t^{s,p}c_xc_yc_z-\mu^{s,p}$, we can plant the Weyl points at ${\bm k}$=(0,0,0) and ($\pm$1,$\pm$1,0), (0,$\pm$1,$\pm$1), and at $(\pm1\pm10)$ momenta for $\mu^{s,p}=-t^{s,p}$ [Fig.~\ref{fig2}(e)]. This vast tunability clearly makes our proposal exciting and applicable for the realization of many nontrivial properties by manipulating a suitable net Chern number in various configurations.

\section{Chiral superconductivity driven Weyl state}\label{Sec:WSC}

The analogy between a spinless $p+ip$-pairing symmetry and 2D Dirac node due to the dual effect of linear-dispersion and definitive chirality is well explored in the literature in different contexts.\cite{TSC,LFu,WeylBalentsSC} We exploit a proximity induced $p+ip$-wave SC between two orbital-selective layers to generate 3D Weyl nodes which are also the pairing nodes. An advantage of a SC triggered Weyl node is that the accompanying particle-hole symmetry $\mathcal{S}$ replaces the spin-rotational symmetry breaking of the former setup. In this case the Bogoliubov-de-Gennes (BdG) fermions transform as $c^n_{{\bm k}\uparrow}\rightarrow \mathcal{S} c^n_{-{\bm k}\downarrow}$ as opposed to $c^n_{{\bm k}\uparrow}\rightarrow \mathcal{R}c^{m\dag}_{{\bm k}\downarrow}$ for the spin-rotation. $\mathcal{S}$ symmetry complements the $\mathcal{M}$ symmetry breaking to preserve the $\mathcal{T}$ symmetry, as in the previous case. To express the BdG Hamiltonian in the same form as Eq.~(\ref{E:Hk}), we shuffle the Nambu operators for the two-orbital model to choose a basis as $\Psi_{\bm k}=(c^s_{{\bm k}\uparrow}, c^{p\dag}_{-{\bm k}\downarrow}, c^p_{{\bm k}\uparrow}, c^{s\dag}_{-{\bm k}\downarrow})$ to obtain
\begin{eqnarray}
H_{\bm k} =  \xi_{\bm k}^+\mathbb{I}\otimes\tau_z+ \xi_{\bm k}^-\Gamma_4 +\Delta(s_{x}\Gamma_1+s_{y}\Gamma_2) -t^{sp}_z s_{z/2}\Gamma_3.
\label{HSC}
\end{eqnarray}
The Hamiltonian implies a SC chiral identity between two orbital selective layers as $\Delta_{\bm k}c^s_{{\bm k}\uparrow}c^s_{-{\bm k}\downarrow}=\Delta_{\bm k}^{\dag}c^{p\dag}_{-{\bm k}\downarrow}c^{p\dag}_{{\bm k}\uparrow}$ where $\Delta_{\bm k}=\Delta(s_{x}+is_{y})$, where $\Delta$ is the amplitude of SC gap. The eigenvalues of Eq.~(\ref{HSC}) come out to be same as that of Eq.~(\ref{E:Hk}), and thus the resulting band dispersion appears to be same as in Fig.~\ref{fig2} for $\Delta=\alpha$ with a single Weyl node at the $\Gamma$ point. Here the Weyl node and SC node coincide on the basal plane, however, they can be split by tuning parameters as before. Interestingly, with suitable parameter choices, one can also engineer a line SC node (`Fermi arc') at the locus of $\xi_{\bm k}^-=s_{x,y}=0$ with various interesting SC properties known to exist for line nodes.\cite{linenode} Evidence and proposals of a chiral $p$ wave are in abundance in condensed matter systems.\cite{He3,Sr2RuO4,CePt3Si,ointerface} Therefore, a proximity induced counter helical SC in adjacent layers is conceivable.

\begin{figure}[top]
\rotatebox[origin=c]{0}{\includegraphics[width=.9\columnwidth]{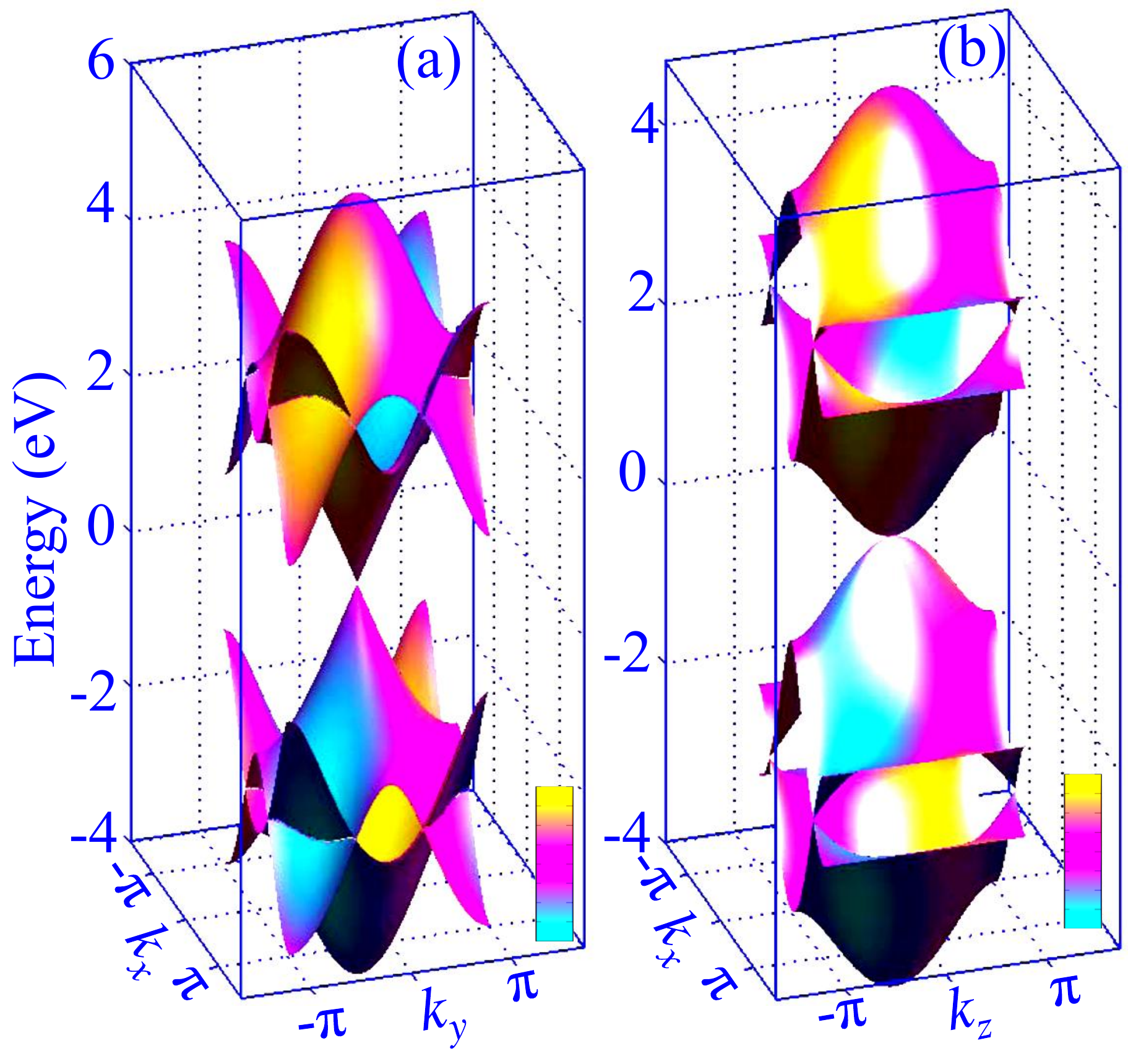}}
\caption{(Color online) (a) A single helical Weyl cone can be produced at the $\Gamma$ point for BdG Hamiltonian (\ref{HSC}) with $\mu^{p}\rightarrow -\mu^{p}$ which gaps out the corresponding $p$ bands. The band dispersion is shown in the $x$-$y$ plane. The $p$ bands exhibit inverted band curvature with a quadratic band bottom and top at the $\Gamma$ point. (b) Same as (a) except that the results are projected here on the $x$-$z$ plane. It should be noticed that the dispersion along the $k_z$ direction becomes quadratic here despite zero bandgap, and the system becomes a semi-Weyl superconductor.}\label{fig3}
\end{figure}

\subsection{Nondegenerate Weyl point}

Elimination of the degeneracy of the Weyl nodes can be achieved for the SC case, by lifting the particle-hole symmetry $\mathcal{S}$ only in one orbital, say the $p$ orbital, while it remains preserved in the $s$ layer. This can be performed by changing the chemical potential $\mu^p\rightarrow -\mu^p$. As a result, the corresponding band becomes inverted and gapped from the nodal region, as shown in Figs.~\ref{fig3}(a) and \ref{fig3}(b) for $k_z=0$ and $k_y=0$ cuts, respectively. This allows us to achieve a long-sought single helical 3D nodal point at the $\Gamma$ momentum, which has been obtained earlier in other classes of topological systems, but not in Weyl semimetals. An important consequence of one helical band at the Fermi energy is the emergence of a Majorana bound state (MBS) which can be achieved here by adding a Zeeman term, and tuning the chemical potential $\mu^s$ to plant the MBS at the zero-bias conductance peak. The stability of a MBS in a proximity induced topological SC (not in a Weyl SC) is studied in Ref.~\onlinecite{proximityMBS}.

\section{Discussion and Conclusions}\label{Sec:Conclusion}

 We propose two complementary design methods for designing 3D Weyl nodal points in an orbital-selective superlattice with the inclusion of either SOC or a spinless $p+ip$-wave SC. The orbital-selective multilayer can be generated by using suitable materials which host opposite orbital symmetries in the low-energy spectrum and at low-dimensions. Rashba-type SOC\cite{BiRashba,PbRashba,SiRashba} and chiral $p$-wave SCs\cite{He3,Sr2RuO4,CePt3Si,ointerface} are readily available in nature. The vast tunability of the parameters offers a desired tool to manipulate the location of Weyl nodes, as well as to obtain a single-helical Weyl band. 

Instead of using only $s$ and $p$ orbitals for the even- and odd parity layers, one can also think of using $d$ and $f$ orbitals, respectively, which will accompany electronic correlations. An electron correlation can be a curse for the stability of Weyl points, however, it can also be a blessing for incorporating several many-body non-trivial phenomena within the Weyl matrix. For example, it has been shown earlier that in the 5$f$ electron URu$_2$Si$_2$, the Fermi surface nesting renders the main band and the shadow bands to touch at the Fermi level (but not at the $\Gamma$ point) with Dirac-like dispersions,\cite{DasHO} which resembles the so-called `accidental degenerate points',\cite{Weyl1} which we call now Weyl points.\cite{Weyl2}   

Given the symmetry of the Hamiltonian deduced here and the opposite parity of the wavefunction in different layers at the topological critical point, i.e., at the Weyl point, strain or other time-reversal invariant perturbations can open an inverted band gap, and the system can be driven from a trivial to nontrivial topological phase. In that case the situation will be analogous to the one proposed earlier in a heterostructure containing an opposite Rashba-type SOC in the adjacent layers.\cite{DasTI} 

In addition to being a noval proposal with possible enhanced tunability for a single or multiple number of Weyl nodes, the present proposal has several advantages: (1) It gives us an opportunity to obtain a single helical Weyl excitation at the $\Gamma$ point. (2) It is easy to tune the net Chern number for the entire momentum space, and thus AHE. (3) The coexistence of Weyl and SC nodes, and line SC nodes, instead of a fully gapped topological superconductor can be explored here. (4) Incorporation of an electron-electron interaction via orbital choices or changing thickness of layers,\cite{DasSODW} is possible. These functional properties can influence the detection and controlled tuning of many non-Abelian topological properties, and electromagnetic responses of our recent interests for conceptual novelties and technologies. (5) Most importantly, the present proposal is free from any crystal geometry restriction, and all symmetry properties and tight-binding parameters for the stabilization of 3D Dirac excitations can be manipulated externally in the heterostructure geometry.

\begin{acknowledgments}
The author thanks A. V. Balatsky, Z. Huang and D. Arovas for valuable discussions. The work is supported by the U.S. DOE through the Office of Science (BES) and the LDRD Program and facilited by NERSC computing allocation.
\end{acknowledgments}


\begin{thebibliography}{10}
\bibitem{footnote} The intraband SOC terms can also be present, however, they do not violate any symmetry operation of our interest, and we have checked that they do not destabilize the Weyl nodes.
\end{thebibliography}
\end{document}